\documentclass[10pt]{article}

\usepackage{graphicx}
\usepackage{indentfirst,csquotes}
\usepackage{amssymb,amsthm,amsmath}
\usepackage{xcolor,paralist,hyperref,titlesec,fancyhdr}
\usepackage{lipsum}
\usepackage{siunitx}    
\usepackage{booktabs}   
\usepackage{float}

\usepackage[
  backend=biber,
  style=numeric,
  sorting=none
]{biblatex}

\newcommand{\br}{\bottomrule}
\newcommand{\mr}{\midrule}

\titleformat{\section}{\normalfont\large\bfseries}{\thesection}{1em}{}
\titlespacing*{\section}{0pt}{1ex}{1ex}

\hypersetup{ colorlinks=true, linkcolor=black, filecolor=black, urlcolor=black }

\begin{document}

\title{Emulation of the Six-State Quantum Key Distribution Protocol with Pulsed Lasers}

\author{
  Sara P. Gandelman$^{1,2}$ \and
  Georgi Gary Rozenman$^{3}$
}

\date{%
  $^{1}$ Raymond and Beverly Sackler School of Physics \& Astronomy, 
  Faculty of Exact Sciences, Tel Aviv University, Tel Aviv 69978, Israel\\
  $^{2}$ School of Electrical Engineering, Iby and Aladar Fleischman Faculty of Engineering, 
  Tel Aviv University, Tel Aviv 69978, Israel\\
  $^{3}$ Department of Mathematics, Massachusetts Institute of Technology, 
  Cambridge, Massachusetts, USA\\[1ex]
  \texttt{garyrozenman@protonmail.com}\\[2ex]
  \today
}

\maketitle

\begin{abstract}
\quad
Quantum cryptography remains a topic of enduring scientific and educational interest. Here, we present a clear and accessible framework for exploring the six-state quantum key distribution protocol—an enhanced, three-basis extension of the BB84 scheme—combining optical experiments with computational analysis. Designed for testing quantum communication protocols through emulation, this approach provides a robust and cost-effective platform that highlights the fundamental principles of multi-basis encoding and demonstrates how experimental measurements connect directly to theoretical expectations in a controlled tabletop setting.
 
\end{abstract} 

\bigskip

\section{Introduction}

Quantum key distribution (QKD) enables two remote parties---conventionally named Alice and Bob---to generate a shared secret key whose security is guaranteed by the laws of quantum mechanics \cite{grasselli2021quantum,mehic2023quantum,portmann2022security}. 
Over an untrusted quantum channel, which may be monitored or even controlled by an eavesdropper (Eve), the protocol allows the legitimate users to detect any interception attempt through the appearance of measurable disturbances~\cite{bb84,wolf2021quantum,lee2022eavesdropping}.

The seminal BB84 protocol achieves this by encoding information in two mutually unbiased bases \cite{rozenman2023quantum}.  
Under a simple intercept--resend (IR) attack, in which Eve measures and retransmits every qubit, she inevitably introduces an error rate of \(25\%\) in the sifted key, providing a clear signature of her presence to Alice and Bob \cite{sun2022review}.  
Building on such foundational schemes, free-space QKD has remained at the forefront of experimental quantum communications, enabling secure key exchange across terrestrial links, airborne platforms, and satellites \cite{sulimany2025high,kam2025reduced}. 
Recent advances in adaptive optics, high-efficiency detectors, and polarization-maintaining optical systems have extended operational distances and key rates, bringing global quantum networks closer to realization \cite{wei2022towards}.  
These developments continue to validate QKD as a leading testbed for quantum information science and applied quantum technologies \cite{lib2025high}.

The six-state protocol enhances this foundational scheme by introducing a third mutually unbiased basis, thereby increasing the symmetry of the qubit space and improving its resilience against eavesdropping~\cite{bruss1998optimal,scarani2008quantum}. 
This extension yields distinct and quantifiable signatures once the basis comparison between sender and receiver is performed. 
For the legitimate exchange between Alice and Bob, only one out of three transmissions corresponds to a matching basis, resulting in approximately \(1/3\) (\(33.3\%\)) of undisturbed bits being retained for key generation. 
When an intercept--resend eavesdropper (Eve) is introduced between Alice and Bob, basis alignment occurs simultaneously among all three parties in only one out of nine cases, leaving just \(1/9\) (\(11.1\%\)) of the bits undisturbed after basis comparison.

True quantum key distribution requires single-photon sources, low-noise detection, and precise optical alignment, conditions that are often impractical in typical university laboratories \cite{hanafi2025quantum}. 
To overcome these limitations, a growing body of work has adopted \emph{classical emulations} of QKD protocols that reproduce the logical sequence and measurement statistics of quantum communication using standard optical components. 
In such experiments, pulsed lasers serve as classical light sources, and polarization optics reproduce the probabilistic behavior of single-photon polarization measurements, allowing students and researchers to explore quantum cryptography principles without specialized quantum hardware~\cite{riggs2022multi}. 

This emulation-based approach has been successfully demonstrated for the BB84 and B92 protocols, providing accessible platforms for studying state preparation, basis selection, and intercept–resend eavesdropping within an educational framework. 
Although these realizations do not achieve true quantum security, they faithfully replicate the statistical outcomes of genuine QKD schemes and thereby serve as powerful pedagogical tools~\cite{bloom2022quantum,gandelman2025hands}. 
More recent implementations extend this concept to emulate higher-dimensional or multi-basis protocols and even entanglement-based systems, combining experimental measurements with computational analysis to bridge abstract quantum theory and hands-on experimentation~\cite{meyer2025analogy,stasiuk2023high,li2023efficient}. 

Within this context, the present work advances the field by providing a six-state QKD emulation---an enhanced, three-basis extension of BB84---that preserves the conceptual and experimental clarity of earlier educational demonstrations while broadening their scope toward multi-basis quantum communication.

In this work, we present a compact, tabletop emulation of the six-state protocol, implemented with a pulsed laser source and bulk polarization optics to reproduce the essential dynamics of quantum key distribution in a controlled classical setting.

We experimentally validate the protocol's security characteristics by comparing its operation in a secure configuration against a simulated intercept--resend attack. 

\begin{figure}[ht]
\centering
\includegraphics[width=\linewidth]{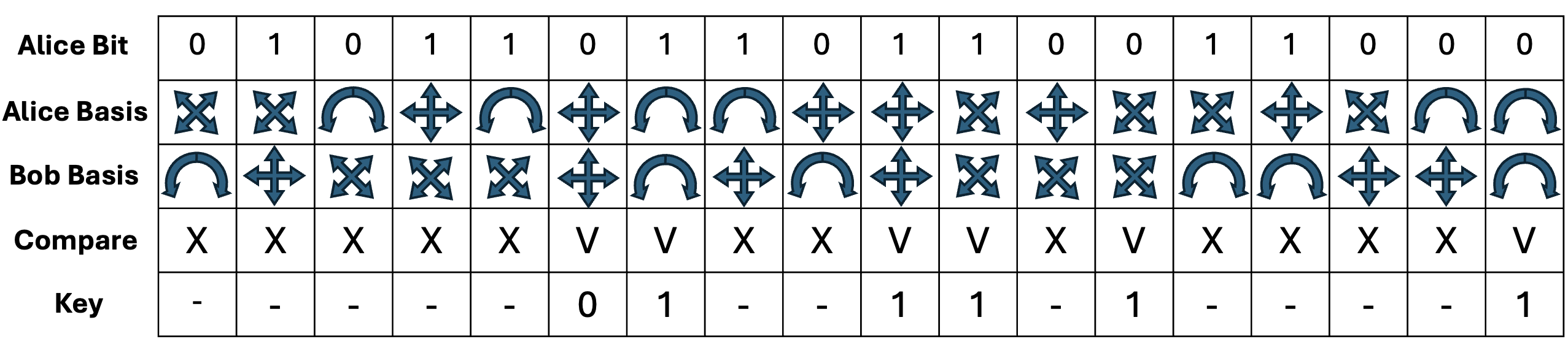}
\caption{\textbf{Core fractions of all transmissions.} 
Right: Alice--Bob (no Eve), undisturbed (sifted) = 1/3 vs.\ disturbed \ 2/3. 
Left: Alice--Eve--Bob (IR), undisturbed-and-compromised  1/9 vs.\ disturbed  8/9. 
Horizontal lines at \(1/3\) and \(1/9\) indicate the theoretical benchmarks.}
\label{fig:core}
\end{figure}

\section{Six-State Protocol and Benchmarks}

The six-state quantum key distribution (QKD) protocol extends the well-known BB84 scheme by introducing a third mutually unbiased basis, as shown in Figure \ref{fig:core}. 
Alice encodes each bit in one of six polarization states \cite{bechmann1999incoherent}:

\begin{equation}
\{\,|H\rangle,\,|V\rangle\,\} \in \mathrm{HV\ basis}.
\label{eq:HVbasis}
\end{equation}

\begin{equation}
|D\rangle = \frac{1}{\sqrt{2}}\!\left(|H\rangle + |V\rangle\right), \quad
|A\rangle = \frac{1}{\sqrt{2}}\!\left(|H\rangle - |V\rangle\right), \quad
\{D,A\} \in \mathrm{DA\ basis}.
\label{eq:DAbasis}
\end{equation}

\begin{equation}
|R\rangle = \frac{1}{\sqrt{2}}\!\left(|H\rangle + i|V\rangle\right), \quad
|L\rangle = \frac{1}{\sqrt{2}}\!\left(|H\rangle - i|V\rangle\right), \quad
\{R,L\} \in \mathrm{RL\ basis}.
\label{eq:RLbasis}
\end{equation}

After transmission, both parties publicly announce their chosen bases and retain only those events where they matched (\(A=B\)).
This process, known as \emph{sifting}, determines which subset of bits contributes to the final raw key \cite{biswas2022modified}.

\vspace{0.5em}
\noindent\textbf{Ideal Sifting Probabilities.}
Since each participant chooses among three bases uniformly at random, the probability that Alice and Bob select the same basis is
\[
P_{\mathrm{sift}} = P(A=B) = \frac{1}{3}.
\]
Thus, in the absence of any disturbance, one third of all transmitted bits are retained as sifted key bits.

\vspace{0.5em}
\noindent\textbf{Intercept–Resend (IR) Benchmark.}
In an intercept--resend (IR) attack, an eavesdropper (Eve) measures each photon in a randomly chosen basis 
\(E \in \{\mathrm{HV},\,\mathrm{DA},\,\mathrm{RL}\}\) and then transmits her measurement outcome to Bob. 
If Eve's basis happens to match Alice's, her result is perfectly correlated with Alice's bit; otherwise, she resends a polarization chosen at random. 
Bob then measures the received photon in his own randomly selected basis. 
The corresponding probabilities are:

\begin{equation}
P(A = B) = \frac{1}{3},
\label{eq:AB_match}
\end{equation}

\begin{equation}
P(E = A) = \frac{1}{3},
\label{eq:EA_match}
\end{equation}

\begin{equation}
P(\mathrm{undisturbed\ and\ compromised}) 
 = P(A = B)\,P(E = A) = \frac{1}{9}.
\label{eq:undisturbed_compromised}
\end{equation}

The \(1/9\) fraction corresponds to the subset of transmissions for which all three bases coincide (\(A=B=E\)); these bits are undisturbed and known to Eve. 
The remaining two thirds of sifted bits are disturbed, manifesting as errors or randomness in Bob’s record. 
Therefore, the expected disturbance (error rate) within the sifted key under IR conditions is \(1/3\) (33.3\%), whereas in a secure, eavesdropper-free channel, the disturbance ideally vanishes.

\vspace{0.5em}
\noindent\textbf{Benchmark Summary.}

\begin{table}[ht]
\centering
\begin{tabular}{lcc}
\br
\textbf{Outcome Type} & \textbf{Alice--Bob (no Eve)} & \textbf{Alice--Eve--Bob (IR)} \\
\mr
Correct (basis matched)   & $1/3$ & $1/9$ \\
Incorrect (basis mismatched) & $2/3$ & $8/9$ \\
\br
\end{tabular}
\caption{Theoretical probabilities for correct and incorrect outcomes in the six-state QKD protocol. 
Without eavesdropping, one-third of all transmissions are correctly matched between Alice and Bob. 
Under an intercept--resend (IR) attack, only one-ninth of all transmissions remain correct, while the remaining fractions are compromised or mismatched.}
\label{tab:benchmarks}
\end{table}

These theoretical benchmarks, shown in Table \ref{tab:benchmarks}, define the expected statistical structure of the six-state protocol and serve as reference values for the experimental results that follow. 
By directly comparing measured fractions of sifted and undisturbed bits against these analytic predictions, one can quantitatively verify the correct operation of the six-state emulation and assess its fidelity relative to the ideal quantum case \cite{liu2022towards}.

\begin{figure}[htbp]
\centering
\includegraphics[width=\linewidth]{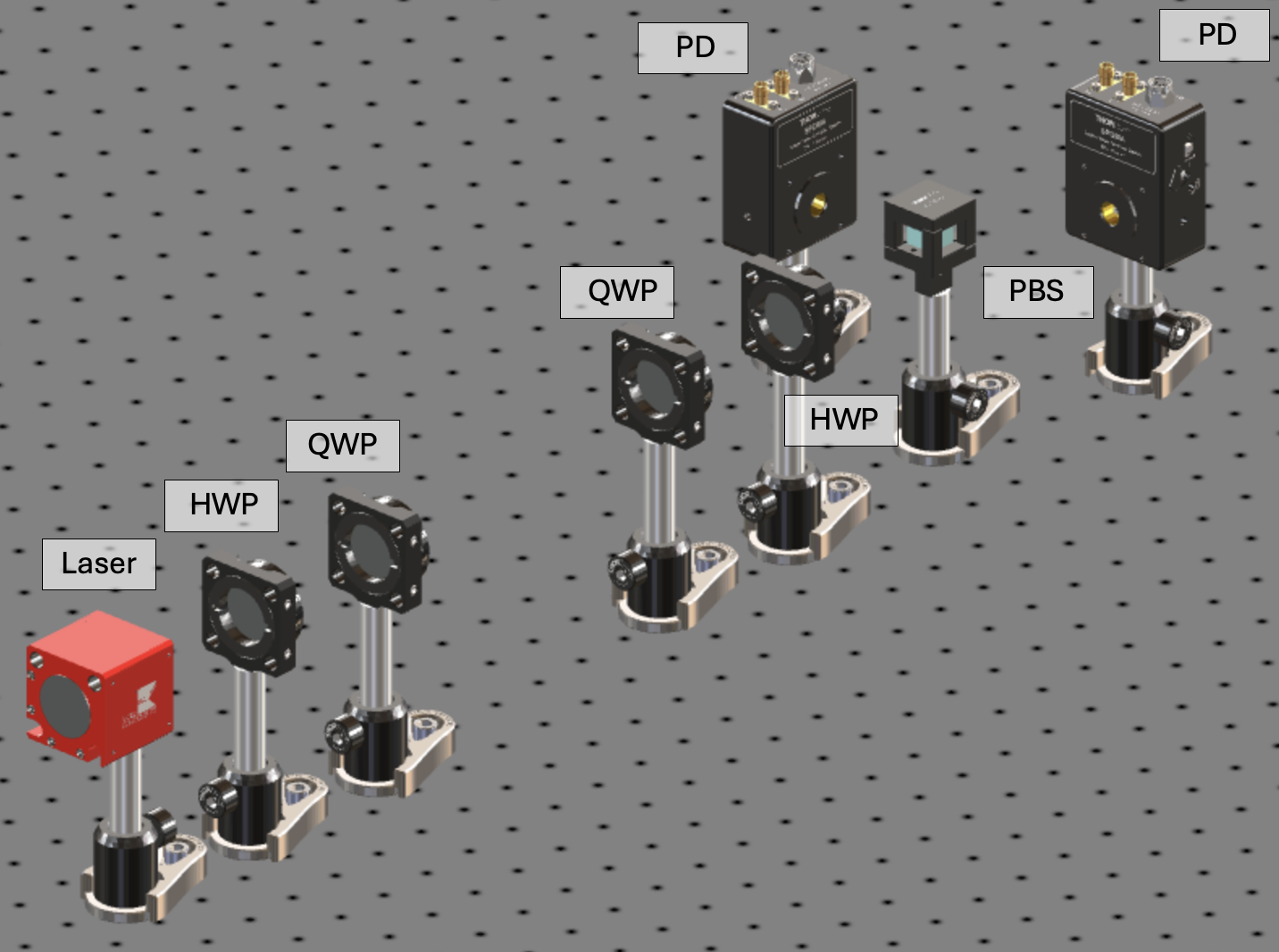}
\caption{\textbf{Experimental setup for the six-state QKD implementation using a pulsed laser source.}
The beam from the laser passes sequentially through Alice’s polarization state preparation stage—comprising a half-wave plate (HWP) and a quarter-wave plate (QWP)—which generates one of the six polarization states \(\{H, V, D, A, R, L\}\). 
The prepared beam is then analyzed by Bob’s measurement module, consisting of a QWP, an HWP, and a polarizing beam splitter (PBS) that projects the incoming state onto the selected measurement basis. The two output ports of the PBS are connected to photodetectors (PD), corresponding to logical bit values 0 and 1.}
\label{fig:core2}
\end{figure}

\begin{figure}[htbp]
\centering
\includegraphics[width=\linewidth]{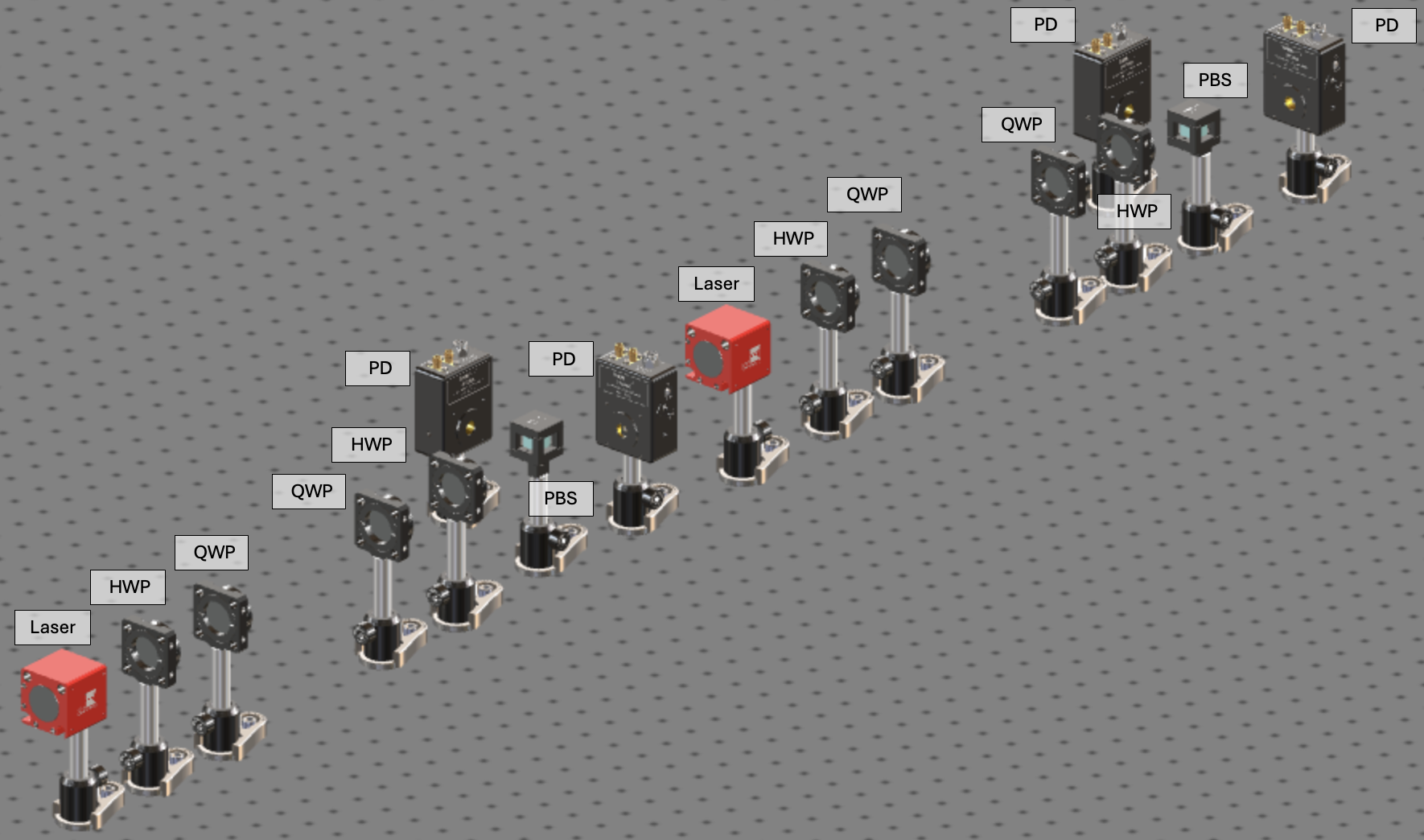}
\caption{\textbf{Experimental setup for the six-state QKD system with an intercept--resend eavesdropper.} 
The optical bench consists of three modules: Alice, Eve, and Bob. 
Alice prepares one of the six polarization states using a laser, half-wave plate (HWP), and quarter-wave plate (QWP). 
Eve intercepts the transmitted beam and performs a full measurement--resend operation using her own QWP, HWP, and polarizing beam splitter (PBS) connected to two photodetectors (PDs). 
A secondary laser, followed by an HWP and QWP, re-encodes the state that Eve resends to Bob. 
Bob’s receiver module (right) analyzes the incoming beam using a QWP, an HWP, and a PBS with two PDs corresponding to logical bit values 0 and 1. }
\label{fig:core3}
\end{figure}

\begin{figure}[htbp]
\centering
\includegraphics[width=\linewidth]{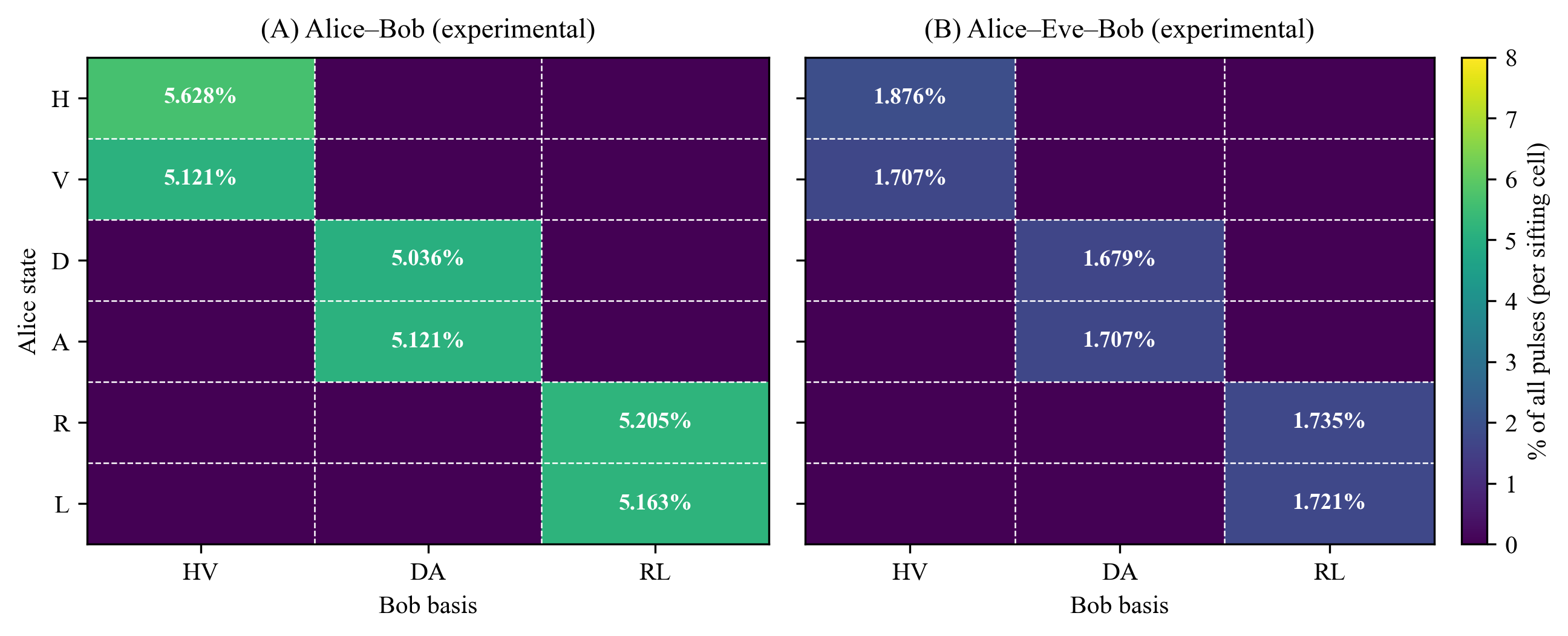}\\[3pt]
\caption{\textbf{Experimental correlation matrices for the six-state QKD protocol.} 
(A) Alice--Bob configuration without eavesdropping, showing detection probabilities for each combination of Alice’s prepared state and Bob’s measurement basis. 
The bright diagonal blocks correspond to correctly correlated basis pairs (\(HV\), \(DA\), and \(RL\)), each exhibiting a mean detection probability of approximately \(5.6\%\) of all transmitted pulses. 
(B) Alice--Eve--Bob configuration under an intercept--resend (IR) attack, where Eve performs a full measurement and resends the state to Bob. 
Here, the correlated regions are reduced to approximately \(1.85\%\) per correlated cell, consistent with the expected theoretical reduction by a factor of three due to the added measurement stage. 
Color intensity represents the percentage of all pulses detected per state--basis combination, as indicated by the color scale on the right.}
\label{fig:masks}
\end{figure}

\section{Experimental Apparatus}

A pulsed diode laser (repetition rate \SI{1}{Hz}, average power \SI{0.1}{mW}) served as a stable classical light source for emulating single-photon transmission. 
The output beam was spatially filtered and collimated before entering Alice’s polarization preparation stage, consisting of a linear polarizer followed by a half-wave plate (HWP) and a quarter-wave plate (QWP). 
By rotating the two wave plates to predefined angular settings, Alice generated the six polarization states of the six-state protocol---horizontal ($H$), vertical ($V$), diagonal ($D$), antidiagonal ($A$), right-circular ($R$), and left-circular ($L$). 
Each prepared state thus corresponds to one of the three mutually unbiased bases (HV, DA, RL).

On the receiver side, as shown in Figure \ref{fig:core2}, Bob’s measurement analyzer mirrored Alice’s polarization control, employing a HWP and QWP combination to select one of the three measurement bases, followed by a polarizing beam splitter (PBS) and two photodiodes (PD1 and PD2) that recorded the transmitted and reflected polarization components. 
The analyzer assembly was mounted on precision rotation stages to ensure reproducible angular alignment across all basis settings.

For the full dataset, a total of $2363$ laser pulses were recorded across all 18 unique state--basis configurations 
(six polarization states prepared by Alice and three measurement bases selected by Bob). 
Each pulse was tagged with the active configuration, the angular settings of the wave plates on both sides, and the binary outcomes of Bob’s photodiodes. 
These data formed the complete raw record used for post-processing and analysis.
The acquisition software logged a configuration label, the angular settings of both wave plates on Alice and Bob, and the binary detection outcomes of the two photodiodes for every laser pulse. 
These raw data streams were later processed to compute per-basis detection probabilities, construct sifting matrices, and evaluate the relative fractions of undisturbed, disturbed, and compromised transmissions.

To emulate an intercept--resend (IR) eavesdropper, we built the experimental apparatus shown in Figure \ref{fig:core3}, the recorded dataset was reprocessed with an additional randomized basis variable representing Eve’s measurement. 
This computational post-processing step simulated Eve’s independent basis choice and re-transmission of each measured polarization, enabling direct comparison between the experimental Alice--Bob case and the simulated Alice--Eve--Bob scenario. 
The resulting analysis reproduces the full logical flow of a six-state QKD exchange while maintaining a purely classical optical implementation, offering a cost-effective and reproducible framework for educational and benchmarking purposes.

\begin{table}[ht]
\centering
\caption{Wave-plate and detector configurations used in the six-state QKD emulation. 
Each row corresponds to one of the 18 combinations of Alice’s prepared state and Bob’s measurement basis. 
Angles are in degrees, and binary entries indicate the logical settings of each wave plate or bit register.}
\label{tab:configs}
\footnotesize
\begin{tabular}{ccccccccccl}
\br
\multicolumn{3}{c}{\textbf{Alice}} & \multicolumn{3}{c}{\textbf{Bob}} & 
\textbf{Alice Bit} & \textbf{Alice Angle} & \textbf{Bob Angle} & \textbf{Config}\\
\cmidrule(r){1-3}\cmidrule(lr){4-6}
HWP1 & HWP2 & QWP & HWP1 & HWP2 & QWP &  & (°) & (°) &  \\
\mr
0 & 1 & 0 & 0 & 0 & 1 & 0 & $-45$ & 0 & A \\
1 & 1 & 0 & 0 & 0 & 0 & 1 & $45$  & 0 & B \\
0 & 0 & 1 & 0 & 1 & 0 & 0 & 0     & 45 & C \\
1 & 0 & 0 & 0 & 1 & 0 & 1 & 90    & 45 & D \\
1 & 0 & 1 & 0 & 1 & 0 & 1 & 90    & 45 & E \\
0 & 0 & 0 & 0 & 0 & 0 & 0 & 0     & 0  & F \\
1 & 0 & 1 & 0 & 0 & 1 & 1 & 90    & 0  & G \\
1 & 0 & 1 & 0 & 0 & 0 & 1 & 90    & 0  & H \\
0 & 0 & 0 & 0 & 0 & 1 & 0 & 0     & 0  & I \\
1 & 0 & 0 & 0 & 0 & 0 & 1 & 90    & 0  & J \\
1 & 1 & 0 & 0 & 1 & 0 & 1 & 45    & 45 & K \\
0 & 0 & 0 & 0 & 1 & 0 & 0 & 0     & 45 & L \\
0 & 1 & 0 & 0 & 1 & 0 & 0 & $-45$ & 45 & M \\
1 & 1 & 0 & 0 & 0 & 1 & 1 & 45    & 0  & N \\
1 & 0 & 0 & 0 & 0 & 1 & 1 & 90    & 0  & O \\
0 & 1 & 0 & 0 & 0 & 0 & 0 & $-45$ & 0  & P \\
0 & 0 & 1 & 0 & 0 & 0 & 0 & 0     & 0  & Q \\
0 & 0 & 1 & 0 & 0 & 1 & 0 & 0     & 0  & R \\
\br
\end{tabular}
\end{table}

\section{Results}
\label{sec:results}
The processed dataset captures all 18 state–basis combinations of the six-state protocol, mapping each configuration to Alice’s prepared polarization state and Bob’s chosen measurement basis. For each configuration, a series of laser pulses was recorded while the polarization optics—namely the half-wave plate (HWP) and quarter-wave plate (QWP)—were randomized according to the settings summarized in Table~\ref{tab:configs}. The detection events were time-tagged and coincidence counts were registered when the signal exceeded a defined signal-to-noise ratio (SNR) threshold, ensuring robust discrimination between true photon events and background noise. Details of the coincidence detection algorithm and SNR calibration procedure are provided in Ref.~\cite{arbel2025optical}

\subsection*{State--basis correlation maps}

Figure~\ref{fig:masks} displays the normalized detection probabilities for each of the 18 state--basis cells, shown separately for the Alice--Bob (A) and Alice--Eve--Bob (B) cases. 
Each cell value represents the percentage of all transmitted pulses that fall within a specific state--basis pairing. 
For the Alice--Bob configuration, the six matched-basis cells (HV, DA, RL) each contribute approximately \(5.5\%\) of all pulses---corresponding closely to the theoretical value of \(1/3\times1/6 \). 
The remaining cells, where the preparation and measurement bases differ, yield negligible correlations, confirming correct optical alignment and basis independence. 

When the intercept--resend eavesdropper is introduced, the undisturbed-and-compromised probabilities per sifting cell decrease to roughly \(1.7\%\), in agreement with the theoretical expectation of \(1/9\times1/6\). 
This reduction appears uniformly across all three matched-basis columns, reflecting that Eve’s random basis choice succeeds only one-third of the time in reproducing Alice’s original state. 
The resulting maps thus visualize, in a direct and quantitative manner, the transition from a fully correlated Alice--Bob channel to a partially disturbed Alice--Eve--Bob channel.

\begin{figure}[htbp]
\centering
\includegraphics[width=\linewidth]{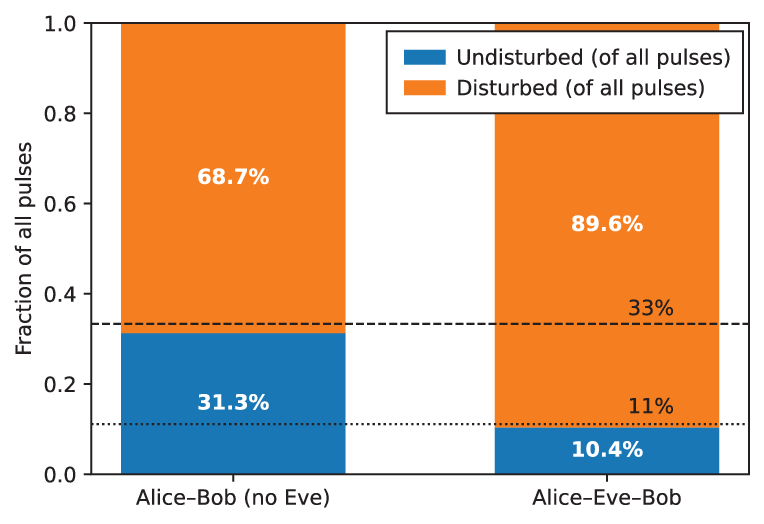}
\caption{\textbf{Core fractions of all transmissions}. Left: Alice--Bob (no Eve), showing an experimentally measured undisturbed (sifted) fraction of approximately \SI{31.3}{\percent}, compared to the theoretical expectation of \(1/3 \approx \SI{33.3}{\percent}\); the remaining \(\approx\SI{68.7}{\percent}\) correspond to disturbed or mismatched events. 
Right: Alice--Eve--Bob (IR) configuration, where the experimentally measured undisturbed-and-compromised fraction is \SI{10.4}{\percent}, close to the theoretical prediction of \(1/9 \approx \SI{11.1}{\percent}\); the remaining \(\approx\SI{89.6}{\percent}\) represent disturbed transmissions. 
Horizontal black dashed and dotted lines mark the theoretical reference levels at \(1/3\) and \(1/9\), respectively, for direct comparison with the experimental data.}
\label{fig:core5}
\end{figure}

Figure~\ref{fig:masks} visualizes the sifting structure. Without Eve, each sifting cell is undisturbed with probability 1; under IR, undisturbed per cell is \(1/3\), producing \(P(\text{undisturbed \& compromised})=P(A=B)P(E=A)=1/9\).

\subsection*{Aggregate fractions and benchmarks}

To emphasize the security-relevant trends in the measured data, the probabilities extracted from Fig.~\ref{fig:masks} are condensed in Fig.~\ref{fig:core5} into two stacked bars, representing the aggregate outcome fractions for both experimental configurations: Alice--Bob and Alice--Eve--Bob. Each bar captures the overall proportion of undisturbed (sifted) versus disturbed or mismatched transmissions, thereby illustrating the protocol’s resilience to measurement disturbance and eavesdropping.

In the secure Alice--Bob configuration (no Eve), the undisturbed or sifted portion accounts for approximately \SI{31.3}{\percent} of all transmitted bits, while the remaining \SI{68.7}{\percent} correspond to disturbed events arising from basis mismatch. This distribution closely matches the theoretical benchmark of \SI{33.3}{\percent} (\(1/3\)), which reflects the probability that Alice and Bob independently select the same measurement basis out of the three available options. The small deviation from the ideal value can be attributed to finite sample size, optical alignment tolerances, and detector noise, all of which are intrinsic to real laboratory conditions.

When an intercept--resend eavesdropper is introduced between Alice and Bob, the undisturbed-and-compromised fraction decreases to \SI{10.4}{\percent}, while \SI{89.6}{\percent} of the transmissions become disturbed. This strong reduction aligns with the theoretical expectation of \SI{11.1}{\percent} (\(1/9\)), corresponding to the probability that all three parties—Alice, Eve, and Bob—happen to choose the same basis simultaneously. The observed agreement confirms that the simulated eavesdropping stage correctly reproduces the logical statistics of the six-state protocol under intercept--resend conditions \cite{yasmin2023modified}.

The horizontal dashed and dotted black lines in Figure~\ref{fig:core5} mark the theoretical reference levels at \(1/3\) and \(1/9\), respectively, providing direct visual benchmarks for the expected sifted and compromised fractions. The close correspondence between these theoretical predictions and the experimental data demonstrates the robustness and accuracy of the optical emulation, validating its use as a reliable and accessible platform for exploring quantum key distribution principles in a classical laboratory environment.

\section{Discussion}

The present experiment demonstrates that classical optical systems can faithfully emulate the statistical behavior of multi-basis quantum communication protocols. By reproducing the key fractions of undisturbed, disturbed, and compromised bits observed in the six-state protocol, the pulsed-laser implementation provides a transparent bridge between theoretical quantum mechanics and hands-on optical experimentation. The close quantitative agreement between the experimental results (\SI{31.3}{\percent} and \SI{10.4}{\percent}) and the theoretical predictions (\SI{33.3}{\percent} and \SI{11.1}{\percent}) validates the emulation approach and confirms that the logical structure of the six-state protocol can be fully explored in a purely classical framework.

Beyond quantum cryptography, emulation represents a universal paradigm across physics \cite{rozenman2019amplitude,gonzalez2019classical,li2021fpga,rozenman2021projectile}. From optical analogs of high-dimensional quantum key distribution to hydrodynamic quantum analogs of particle-wave duality, and from laboratory models of gravitational systems to photonic simulations of quantum chaos, emulation serves as a unifying experimental methodology \cite{rozenman2024observation,marvian2025efficient,mourya2023emulation,diadamo2021integrating}. In all such cases, controlled classical platforms reproduce key mathematical features of their quantum or relativistic counterparts—providing physical intuition, accessible experimentation, and opportunities for cross-disciplinary education \cite{how2022advancing}.

\section*{Acknowledgements}

{We thank \textsc{3D-Optix} for providing the license for the optical design software used to visualize the experimental setup. We gratefully acknowledge the \textit{Raymond and Beverly Sackler School of Physics and Astronomy} at Tel Aviv University for access to laboratory facilities and instrumentation. We also thank Prof. Ady Arie, Prof.~Haim~Suchowski and Prof.~Shimshon~Bar-Ad for their continued support. G. G. R.~gratefully acknowledges the generous support of the \textit{C.~L.~E.~Moore Instructorship} in the Department of Mathematics at the Massachusetts Institute of Technology.
}

\section*{Data Availability Statement}

Data sets generated during the current study are available from the corresponding author on reasonable request.

$\,$

$\,$

\printbibliography

\end{document}